**Optical characterization of deep level defects in n-type $Al_xIn_yGa_{1-x-y}P$ for development of solid-state photomultiplier analogs**

**Andrew M. Armstrong, Evan M. Anderson*, Lisa N. Caravello, Eduardo Garcia, Joseph P. Klesko, Samuel D. Hawkins, Eric A. Shaner, John F. Klem, Aaron J. Muhowski**

***Corresponding Author: E-mail: emander@sandia.gov, aarmstr@sandia.gov and ajmuhow@sandia.gov**

**Abstract**

Characterizing intrinsic defects is an important step in evaluating materials for new optoelectronic device applications. For photomultipliers, suppressing dark currents is critical, but there exists a tradeoff between maximizing the band gap while remaining sensitive to the wavelength of interest, and minimizing the incorporation of new defects by growing not-yet-optimized alloys. We present a series of capacitance-based measurements, including deep level optical spectroscopy, steady-state photocapacitance and illuminated capacitance-voltage, on photodiodes with lightly *n*-type $Al_xIn_yGa_{1-x-y}P$ absorber regions. Several deep levels are identified, including one near midgap. While the inclusion of aluminum increases each trap density by approximately 10x, the hole capture cross section also appears to decrease, suggesting that Shockley-Read-Hall dark currents may be suppressed. These materials may be good candidates for development into silicon photomultiplier analogs with wider bandgap for scintillator applications.

## 1. Introduction

Silicon photomultipliers (SiPMs) have been widely studied as a replacement for photomultiplier tubes (PMTs) in scintillation-based detection architectures [1–9]. Unlike vacuum tube based PMTs, SiPMs are highly manufacturable, compact devices that benefit from decades of investment into silicon semiconductor fabrication processes. Despite these advantages, PMTs exhibit orders of magnitude lower dark count rates than SiPMs, which, as semiconductor devices, have additional dark current mechanisms that can trigger dark counts, such as generation-recombination (GR) and diffusion dark current. In order for SiPMs or SiPM-like devices to compete with PMTs in high performance scintillators, such dark currents must be significantly reduced.

SiPMs comprise an array of Geiger-mode avalanche photodiodes; the dark current processes in an individual photodiode depend exponentially on the bandgap of the constituent semiconductor. Diffusion dark current dominates in an ideal photodiode and can be expressed as

$$J_{diff} = \frac{q\, n_i^2 w}{n_0 \tau_{mc}}$$

for $q$ the elementary charge, $n_i$ the intrinsic carrier density, $w$ the lesser of either the minority carrier diffusion length or the width of the absorber region, $n_0$ the majority carrier density, and $\tau_{mc}$ the minority carrier lifetime. In materials with short Shockley-Read-Hall lifetimes, the contribution from GR dark current can also be significant. The GR dark current density can be expressed as

$$J_{GR} = \frac{q\, n_i w_{dep}}{\tau_{mc}}$$

where $w_{dep}$ is the width of the depletion layer. The intrinsic carrier density, which appears in both expressions, in turn depends on the bandgap ($E_g$) of the absorber material as

$$n_i \propto \exp\left(-\frac{E_g}{2\, k_B T}\right) .$$

Ideally, the largest possible bandgap that maintains appreciable optical absorption at the scintillator wavelength should be chosen to suppress diffusion and GR dark current processes. However, such materials with smaller bandgaps may still be optimal if the minority carrier lifetime is significantly longer than another candidate material with a larger bandgap.

Semiconductor systems offer variability in the available bandgaps and technological maturity, both of which must be considered when identifying candidates for next-generation SiPM-like detectors. Consider the fluorescence photon energy from commonly-used NaI(Tl) scintillators of 3 eV, for example, which precludes the use of wide-bandgap materials like GaN and 4H-SiC, as they would not significantly absorb photons from the scintillator crystal [10,11]. Among III-V materials, the alloy $Al_xGa_{1-x}P$ has the widest bandgap that could mediate photodetection in a scintillator, however available GaP substrates are small, expensive, and burdened by high etch pit densities [12]. The $Al_xIn_yGa_{1-x-y}P$ system lattice-matched to widely-available and mature GaAs substrates can provide nearly the same bandgap with high Al content, ranging from 1.9 to 2.4 eV—approximately twice that of Si. While the increased bandgap suggests dark processes may be reduced, such a reduction still requires that the minority carrier lifetimes be sufficiently long. General trends within III-V semiconductors suggest that the inclusion of Al incurs an increase in oxygen incorporation and consequent point defect densities. At present, there is limited knowledge of minority carrier lifetimes, much less information on deep levels that contribute to Shockley-Read-Hall recombination, that are attainable in low-doped $Al_xIn_yGa_{1-x-y}P$ candidate materials [13–15]. Nevertheless, avalanche photodiodes composed of InGaP and InAlP and even SiPM-like devices have demonstrated promising performance, suggesting that further studies of $Al_xIn_yGa_{1-x-y}P$ are warranted [16–18].

Here, we present a deep level optical spectroscopy (DLOS) study of photodiodes comprising $In_{0.49}Ga_{0.51}P$ (InGaP hereafter) and $Al_{0.13}In_{0.48}Ga_{0.39}P$ (AlInGaP hereafter) absorber materials. Previous studies of deep level defects in InGaP- or AlInGaP-based solar cells or light emitting diodes used deep level transient spectroscopy (DLTS) [19–23]. However, DLTS is a thermally-stimulated technique that is typically limited to observing deep levels that lie < 1 eV from a band edge and thus might not be able to observe near-mid-gap deep levels in $Al_xIn_yGa_{1-x-y}P$ that could produce dark current and limit minority carrier lifetime. As an optically-stimulated technique, DLOS is able to observe deep levels lying near-mid-gap in $Al_xIn_yGa_{1-x-y}P$. DLOS has been performed on InGaP- [24] and AlInGaP- [25] based solar cells, but III-P-based solar cells use *p*-type absorber regions, whereas the photodiodes in this study use *n*-type absorber layers. Deep level defect incorporation is expected to differ for *n*- versus *p*-type AlInGaP alloys because the formation energy of native defects can depend strongly on the energetic position of the Fermi level. Thus, our DLOS study focuses on lightly doped *n*-type materials which could be used in *p*-*v*-*n* avalanche photodiodes. We identify three deep level defects in both InGaP and AlInGaP, two of which have similar distances from the band edges and thermodynamic energies that suggest analogous atomistic origins of these defects for the two materials. We also observe the deep level density increasing by more than an order of magnitude with the inclusion of Al. Despite the high trap density, our study suggests that the hole capture cross section may be significantly smaller in AlInGaP than in InGaP, potentially yielding a minority carrier lifetime, and thus dark current, advantage for AlInGaP over InGaP.

**2. Methods**

Samples were grown by molecular beam epitaxy using a phosphine gas source and a solid-source arsenic cracker on *n*-type, (100) GaAs substrates. *n*- and *p*-type dopants were Si and Be, respectively. For both the InGaP and AlInGaP absorber materials under study here (Fig. 1), growth began with a 100 nm *n*-GaAs buffer layer prior to lowering the growth temperature to 515 °C for 100 nm of *n*-InGaP doped to $1 \times 10^{18}$ cm$^{-3}$ followed by 900 nm of the absorber material. The first 100 nm were doped $1 \times 10^{18}$ cm$^{-3}$ *n*-type, the next 700 nm $6 \times 10^{16}$ cm$^{-3}$ *n*-type, and the final 100 nm $1 \times 10^{18}$ cm$^{-3}$ *p*-type. From there, a 23 nm digital grade was grown to a *p*-type InAlP window layer at a doping of $1 \times 10^{18}$ cm$^{-3}$. Finally, a heavily *p*-type InGaP and GaAs contact structure was grown.

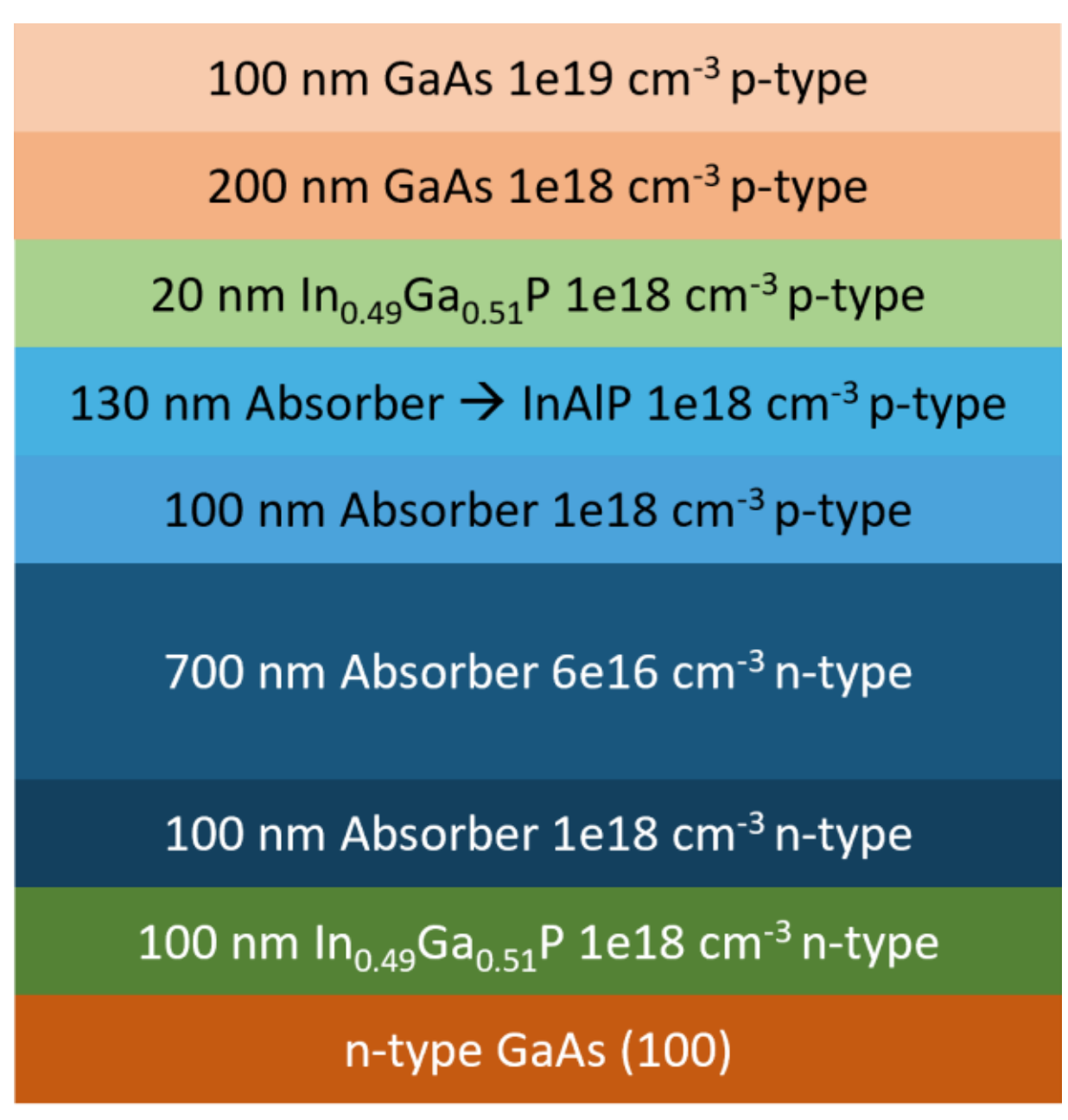

Figure 1. Schematic cross section of the epitaxial samples, where the absorber is either $In_{0.49}Ga_{0.51}P$ or $Al_{0.13}In_{0.48}Ga_{0.39}P$.

Photodiodes were fabricated using standard photolithographic techniques (Fig. 2). First, mesas were reticulated by wet chemical etching through the arsenide and phosphide layers to the underlying substrate. The aqueous HCl or HCl:$H_3PO_4$ chemistry that is typically used to selectively etch phosphides over arsenides laterally etches Al-rich phosphides, namely the InAlP layer, much faster in <010> directions than <011> directions and severely undercuts the mask [26].

This led to development of a non-selective etch using aqueous $HNO_3$:HCl to balance the slow etch rate with low HCl concentration versus the anisotropic etching with high HCl concentration. The devices were then encapsulated in approximately 2000 Å of $Si_3N_4$ to provide electrical isolation between the semiconductor and offset bond pads. Next, vias were etched through the $Si_3N_4$ via reactive ion etching to allow for ohmic contacts to the top of each diode. The top electrical contacts were patterned as an annulus to allow for an optical window for measurements. The metals for the *p*-type (top of diode) and *n*-type (back of wafer) electrical contacts, consisting of at least 20 nm of an adhesion layer and 200 nm of Au, were then deposited on the devices, followed by a 30 s rapid thermal anneal at 400 °C under argon flow. Finally, the GaAs layer was etched from the optical windows using the metal and remaining $Si_3N_4$ as a mask to prevent the GaAs from interfering with the optical measurements of the underlying phosphide layers.

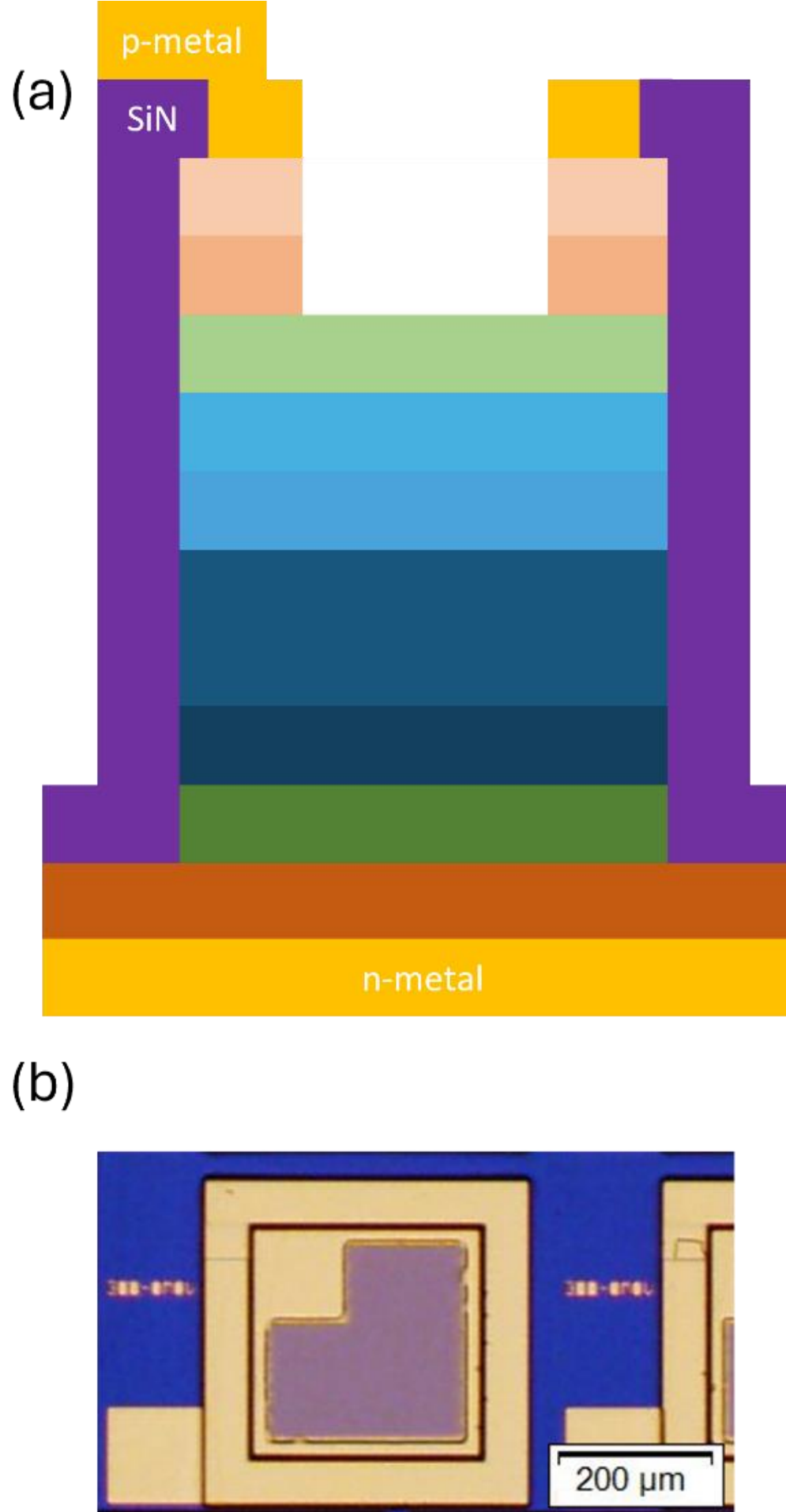

Figure 2. (a) Cross sectional schematic of the fabricated photodiodes. (b) Optical image of a fabricated diode.

The photodiodes were characterized using dark and lighted current-voltage (*I*-*V*) and capacitance-voltage (*C*-*V*). All *C*-*V* and *I*-*V* measurements were performed at room temperature. The doping level in the absorber regions near the p-n junction was measured by *C*-*V*. The *C*-*V* data for both types of photodiodes, shown in Fig. 3, were collected using a 1 MHz frequency with a 30 mV ac amplitude. To extract the net doping ($n_0$) from *C*-*V* measurements, the physical mesa area ($A$) was used as the junction area for the diodes. Uncertainty in $n_0$ can be determined from standard error propagation analysis of the usual equation relating $n_0$ and $C(V)$:

$$n_0 = \frac{C^3}{q\epsilon A^2 \frac{dC}{dV}}$$

where $A$ is the junction area, $q$ is the fundamental charge and $\epsilon$ is the dielectric permititivty in semiconductor. The square mesa sidewall length (nominally 300 $\mu$m) variation was assumed to be 2 $\mu$m due to photolithographic variations. This sidewall length uncertainty results in 0.9% uncertainty ($\sigma_A/A$). Mesa area uncertainty was expected to be the largest source of experimental error for extracting doping from $C$-$V$ measurements because the capacitance meter resolution was small (1 fF) compared to the junction capacitance (19 – 29 pF, depending on bias) and the voltage resolution was small (0.001 V) compared to the applied voltage (~ -5 V), implying percentage errors of $C^3$ and $dC/dV$ << 1%. The extracted doping for a given $C(V)$ value then varied as $n_0 \sim k/A^2$, where $k$ is a constant with negligible uncertainty compared to $A$, so the percentage error of $n_0$ ($\sigma_{n0}/n_0$) ~ $(2\sigma_A/A)^{1/2}$ = 1.3%. The extracted $n_0$ was 6.4 × $10^{16}$ $cm^{-3}$ for the AlInGaP photodiode, and $n_0$ = 3.4 × $10^{16}$ $cm^{-3}$ for the InGaP photodiode. As explained in the Results and Discussion section below, lower doping in the InGaP photodiode resulted from lower-than-expected dopant incorporation rather than high levels of dopant compensation by defects.

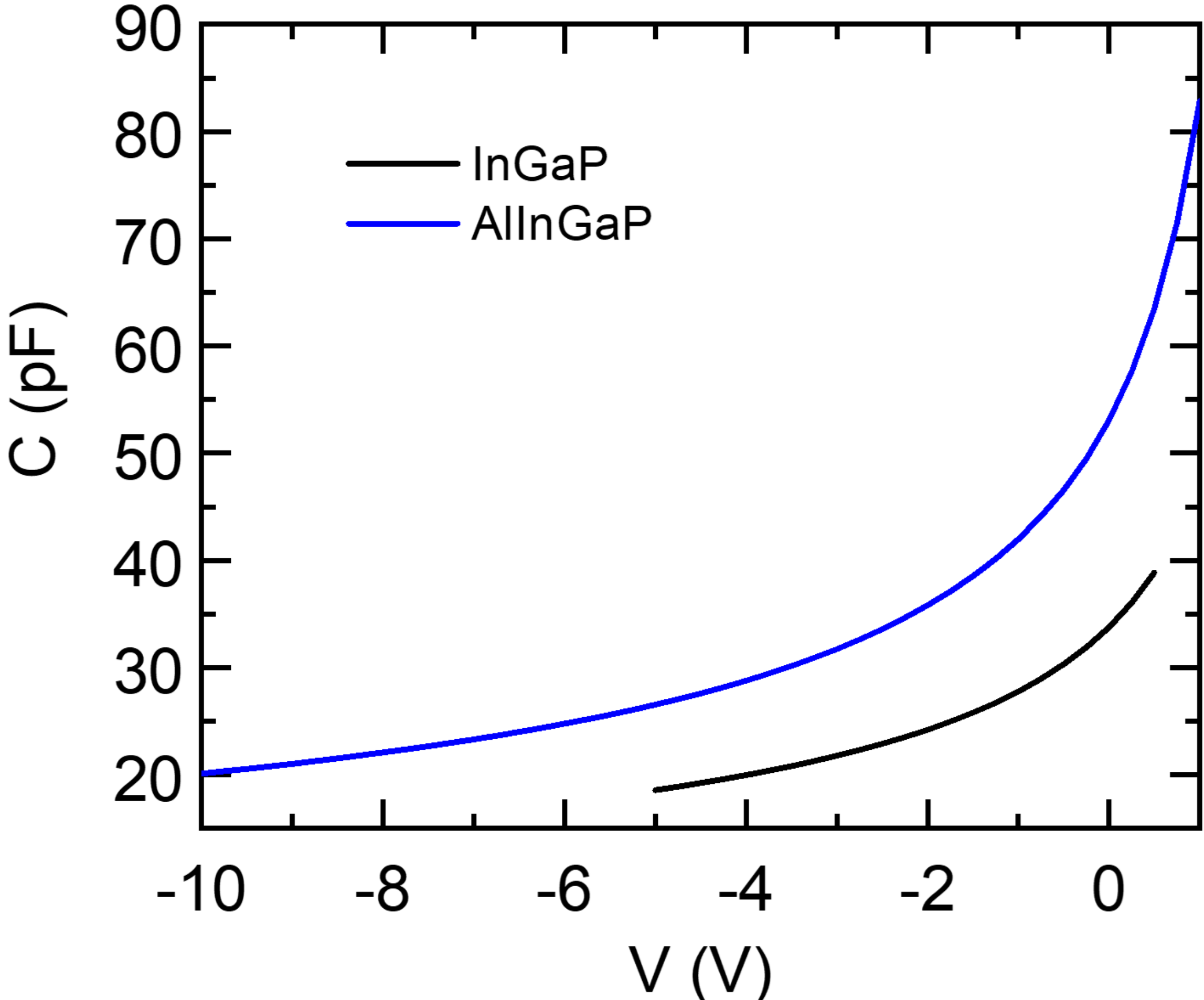

Figure 3. *C-V* data for the photodiodes.

DLOS, steady-state photocapacitance (SSPC) and lighted capacitance-voltage (LCV) were used to study and compare deep level defect states for the two different alloys. DLOS, SSPC and LCV measurements were performed at room temperature. DLOS determines the optical deep level energy ($E^o$) relative to the majority carrier band edge (the conduction band in this case), while SSPC measures the deep level concentration ($N_t$) when $N_t$ is uniform and much lower than the dopant concentration ($N_d$). For the cases when $N_t \sim N_d$ or when $N_t$ has a strong depth-dependence, LCV can be used to measure $N_t$ [27,28].

DLOS is a differential photocapacitance technique that uses monochromatic, sub-bandgap energy photons to measure the deep level optical cross-section ($\sigma^o$) [29]. Fitting the line-shape of $\sigma^o$ to a model [30,31] determines $E^o$ and the Franck-Condon energy ($d_{FC}$). The authors chose the model of Pässler [31] because it is more parsimonious compared to the model of Chantre et al. [29]. The Chantre model includes an ad-hoc "*m*" fitting parameter that accounts for the degree of admixture among conduction- and valence-band-like states for the deep level defect wave function, but this parameter has no direct, physical analog. Conversely, the Pässler model only includes physical values of optical deep level energy, Franck-Condon energy, and local vibrational phonon energy.

The optical deep level energy is the minimum $h\nu$ required to delocalize a carrier from a deep level defect into an energy band in the absence of any phononic interaction between the defect and the lattice. The Franck-Condon energy is the average vibronic energy of a localized carrier on a defect at a given temperature due to phonon excitation of local breathing modes. The photon energy required to delocalize a carrier from a defect is reduced by $d_{FC}$, so the average $h\nu$ required for photoemission of a localized carrier from a defect is $E^o - d_{FC}$. At finite temperature, defects in the lattice have an ensemble of vibronic energies, which gives rise to homogenous broadening of $h\nu$ required to induce photoexcitation from a deep level. This manifests as a temperature-dependent broadening of the defect absorption spectrum, i.e. $\sigma^o$. For the case of strong lattice coupling by the defect, i.e. strong electron-phonon interaction of a carrier localized on the defect, $E^o$ and $d_{FC}$ values can be extracted by fitting $\sigma^o$ to the model of Pässler [31], which uses an amplitude pre-factor, the average defect breathing mode energy ($\varepsilon$), $E^o$ and $d_{FC}$ as variables. For the case of weak lattice coupling by the defect, the resulting sharp $\sigma^o$ can be described by the model of Lucovsky [30] that uses only an amplitude pre-factor and $E^o$ as variables. The model of Pässler reduces to the Lucovsky model for the case of $d_{FC} = 0$.

Deep level $\sigma^o$ was determined experimentally by measuring photocapacitance transients $\Delta C(t)$ arising from deep level defect photoemission upon exposure to monochromatic photons. For electron photoemission from deep level defects located in an *n*-type depletion region, $\Delta C(t)$ is an exponential decay characterized by an optical emission rate [32] $e^{o,n}(h\nu) = \sigma^o(h\nu)\phi(h\nu)$, where $\phi$ is the incident photon flux. Thus, $\sigma^o$ is determined by measuring $\Delta C(t)$ at several $h\nu$ values and extracting $e^{o,n}$ from a least-squares fit [an example of $\Delta C(t)$ data and their fitting is shown in Fig. S1 in the Supplementary Material] and then normalizing to $\phi$. For positive (negative) $\Delta C$, $E^o$ is referenced to the majority (minority) carrier band edge. The experiment proceeded as follows. The diode was held in the dark at room temperature under a reverse measurement bias ($V_r$), and it was assumed that all deep level defects in the depletion region were fully occupied. Monochromatic light was provided using a Xe arc lamp source filtered through a 1/4 meter monochromator with mode-sorting filters to achieve monochromatic illumination with $\phi = 1 \times 10^{17}$ cm$^{-2}$s$^{-1}$ that was held constant by adjusting the monochromator slit width. Variable slit width resulted in a monochromator energy resolution of ~ 5 meV for $0.6 < h\nu < 1.5$ eV and 15 – 25 meV for $1.5 < h\nu < 2.4$ eV. A collimating lens and a focusing lens were used to produce an output beam from the monochromator that was approximately 1 mm × 3 mm. Photon flux was measured for each $h\nu$ using an optical power meter and accounting for the beam size. Diodes were illuminated through the top-side annular metal contact. DLOS was measured at a $V_r$ = -5 V. After recording $\Delta C$(t), an electrical fill pulse bias ($V_f$) = +1 V was applied to reduce the depletion region width and allow deep level defects to thermally re-capture carriers. The positive $V_f$ was below the diode turn-on voltage but still produced several hundred microamps of forward current that provided low-level hole injection into the $n$-regions of the $p^+$-$n$ diode. A thirty second delay between the end of the electrical fill pulse and beginning of subsequent illumination allowed time for any thermal capacitance transient to decay prior to optical excitation of deep level defects.

The above analysis relating $\sigma^o$ and $e^o$ is valid for photoemission of a single carrier type (electron or hole). However, for $h\nu$ greater than half the band gap energy, simultaneous optical emission of both electrons and holes from a deep level can occur, giving $e^o = (e^{o,n} + e^{o,p})$ for $\Delta C(t)$. Determining $E^o$ requires separating $e^{o,n}$ and $e^{o,p}$. Since the steady-state photocapacitance $\Delta C_{ss} \propto e^{o,n}/(e^{o,n} + e^{o,p})$, $E^o$ can be determined from $\sigma^{o,n} = e^{o,n}/\phi \propto e^o \Delta C_{ss}/\phi$ [2].

SSPC and LCV are methods to determine $N_t$. Using SSPC, $N_t = 2n_0 \Delta C_{ss}/C_0$, where $C_0$ is the capacitance in the dark. However, this expression for $N_t$ only holds for the case of small $\Delta C/C_0$, i.e., small $N_t/n_0$. For the AlInGaP device, $\Delta C$ was comparable to $C_0$ and thus SSPC could not be used to measure $N_t$, so LCV was used instead. LCV measures $N_t$ from the difference of space-charge density profiles measured by $C$-$V$ under sub-band gap illumination. Photon energies are chosen to selectively depopulate deep level defects one at a time, and the increase in space-charge density determined from $C$-$V$ provides $N_t$ as a function of depth. For LCV, $N_t$ had an uncertainty

$$\sigma_{N_t}(x_d) = \sqrt{[(\sigma_{n0}/n_0) * n_{h\nu 1}(x_d)]^2 \; + \; [(\sigma_{n0}/n_0) * n_{h\nu 2}(x_d)]^2}$$

where $x_d$ is the depletion depth, $n_{h\nu 1(2)}$ refers to the extracted doping value for photon energies $h\nu_1$ or $h\nu_2$ and the percentage uncertainty in $n_{h\nu 1(2)} = \sigma_{n0}/n_0$ for that of $C$-$V$ measurements, as described above. Evaluation of $\sigma_{Nt}$ is provided in the Results and Discussion section below.

## 3. Results and Discussion

Figure 4 shows the DLOS spectrum for the InGaP diode. Two deep levels with broad spectral line-shapes were observed, and simultaneous fitting of both to the model of Pässler [31] determined $E^o$ = 1.06 eV ($d_{FC}$ = 0.28 eV) and $E^o$ = 1.58 ($d_{FC}$ = 0.16 eV), respectively. A third deep level with a sharp line-shape was fit using the Lucovsky model [30] to determine $E^o$ = 1.81 eV. Non-linear least-squares regression fitting of the DLOS data in Fig. 4 to the model of Pässler [31] produced a root mean squared error (RMSE) of 0.05 eV. This value was taken to be the uncertainty for both $E^o$ and $d_{FC}$ values for the $E_c$ – 1.06 eV and $E_c$ – 1.58 eV deep levels. Non-linear least-squares regression fitting of the DLOS data in Fig. 4 to the model of Lucovsky produced RMSE = 0.07 eV, which was taken to be the uncertainty for $E^o$ for the $E_c$ – 1.81 eV deep level. The DLOS spectrum saturates at the InGaP band gap energy of 1.90 eV because free carrier absorption, which creates a photocurrent rather than photocapacitance, begins to dominate. As shown below, $\Delta C$ for these deep levels were positive, indicating majority carrier emission to the conduction band, so their $E^o$ values are referenced to the conduction band minimum energy ($E_c$).

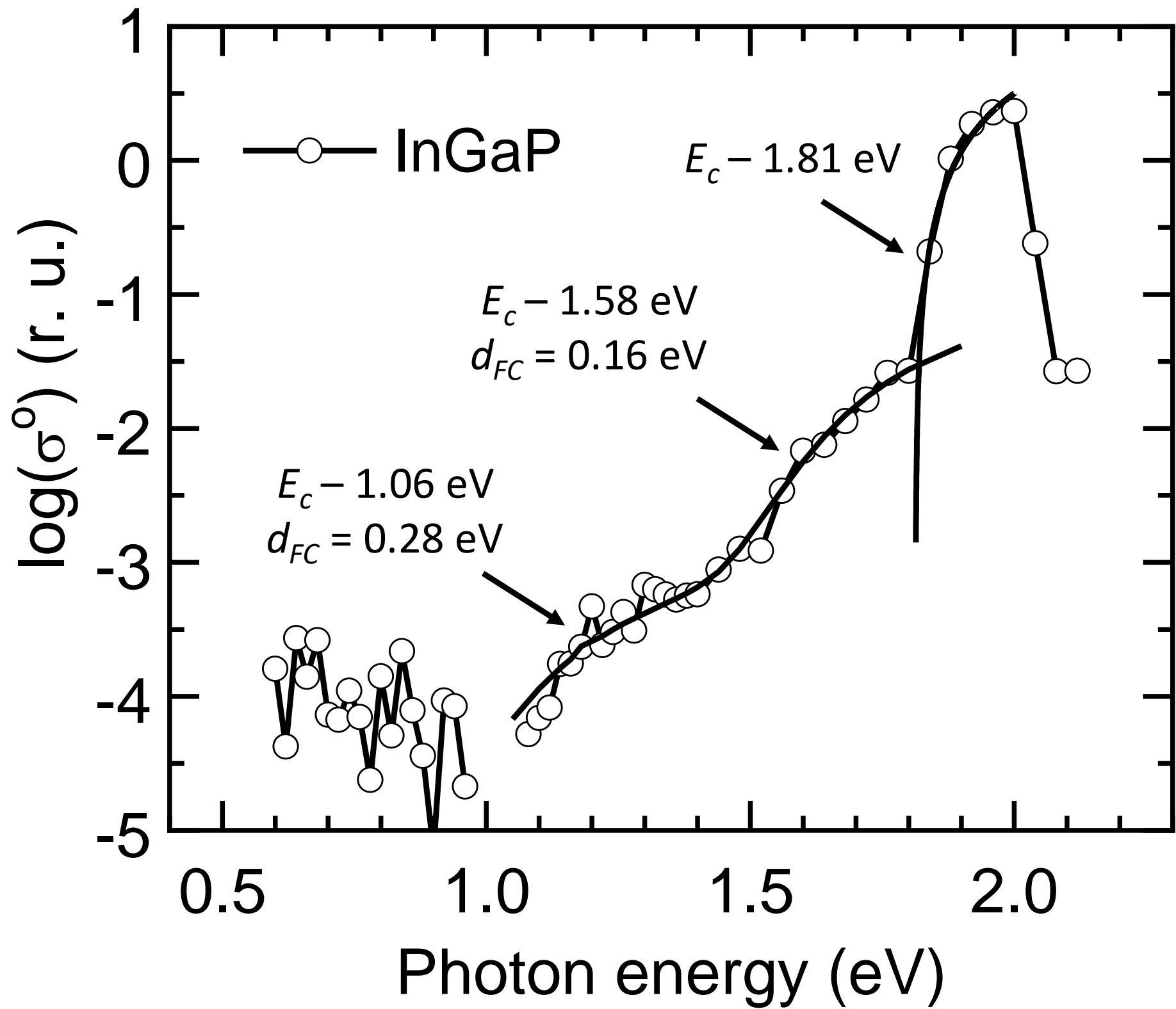

Figure 4. DLOS spectrum of the InGaP diode. The symbols are experimental data and the curves are fits to the Pässler model [31] for the $E_c$ – 1.06 eV and $E_c$ – 1.58 eV levels and the Lucovsky [30] model for the $E_c$ – 1.81 eV level. The scatter in the data for $h\nu$ < 1 eV indicates the noise floor of the measurement where no deep levels were observed.

SSPC determined the concentration of the deep levels observed by DLOS and revealed the role of the $E_c$ – 1.06 eV deep level as a recombination center. The SSPC spectra in Fig. 5 display three positive changes in slope at 1.15, 1.55 and 1.80 eV, correlating to the onset of electron photoemission from the $E_c$ – 1.06 eV, $E_c$ – 1.58 eV, and $E_c$ – 1.81 eV deep levels, respectively. For each deep level, $\Delta C_{ss}/C_0$ was small, so Fig. 5 shows the corresponding $N_t$ data. SSPC is a cumulative measurement, so $N_t$ of the $E_c$ – 1.58 eV deep level is the change in net $N_t$ from 1.55 – 1.80 eV. Likewise, $N_t$ for the $E_c$ – 1.81 eV deep level is the change in net $N_t$ from 1.80 – 1.90 eV. Ascription of the SSPC signal between 1.80 – 1.90 eV to the $E_c$ – 1.81 eV agrees with the abrupt emergence of this deep level at 1.80 eV in the $\sigma^o$ spectrum of Fig. 4. Partitioning the SSPC signal from the $E_c$ – 1.58 eV level with broader $\sigma^o$ is ambiguous because it lacks an abrupt absorption threshold in Fig. 4. In this case, 1.55 eV was chosen as the demarcation in the SSPC signal onset for the $E_c$ – 1.58 eV deep level based on deviation from a horizontal line guide to the eye. Note that the SSPC signal near 1.50 eV is slowly varying, so assignment of SSPC demarcation energy in this region does not have significant impact on $N_t$.

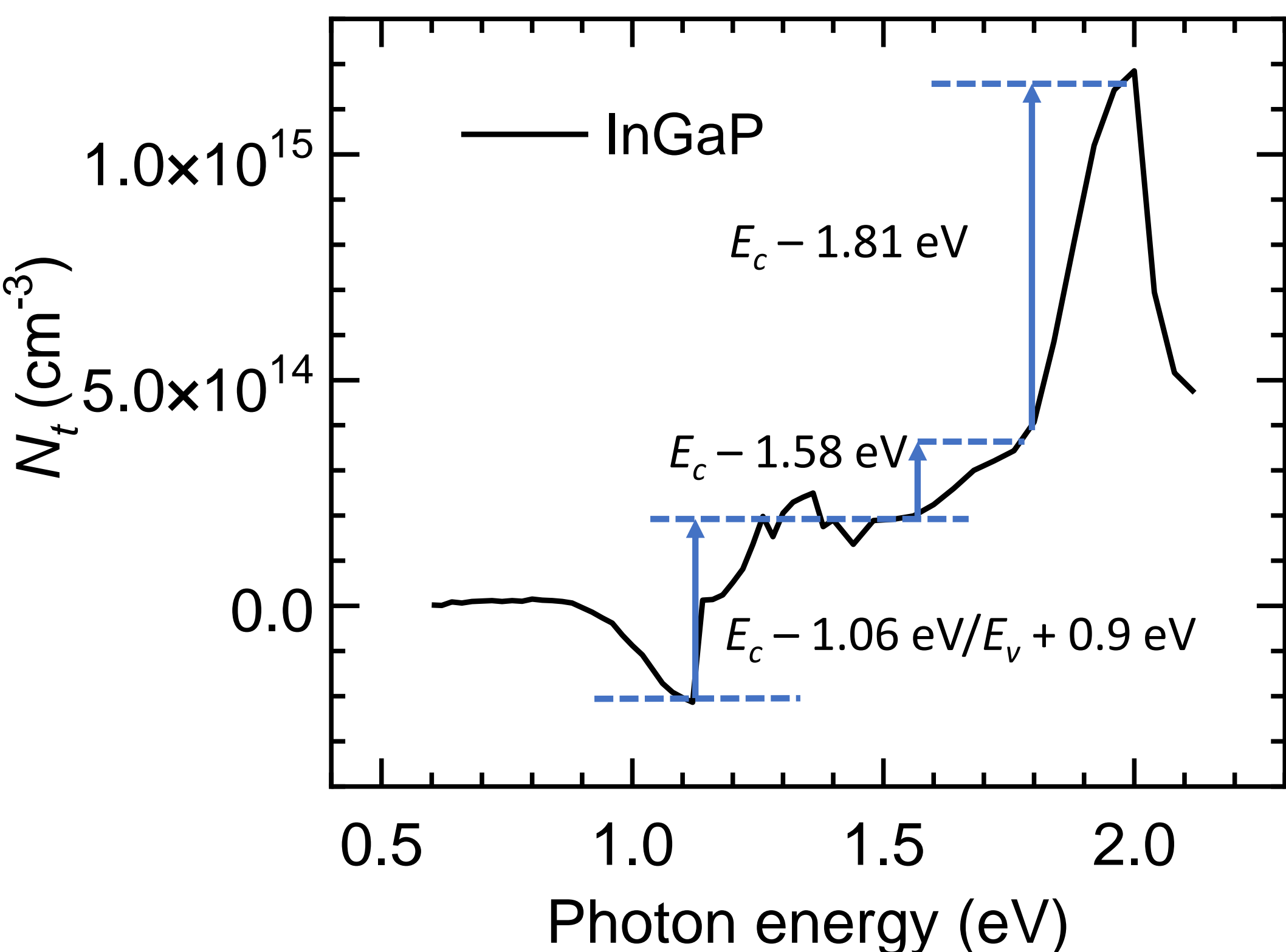

Figure 5. SSPC spectrum of the InGaP diode. The arrows indicate how the total deep level concentration was apportioned among the individual deep levels.

Analysis of $N_t$ for the $E_c$ – 1.06 eV deep level is more involved because its positive $\Delta C_{ss}$ is preceded by a negative threshold at 0.9 eV. Negative SSPC indicates minority carrier photoemission, in this case holes to the valence band maximum $E_v$. Thus, the physical interpretation of negative $N_t$ in Fig. 5 is the concentration of holes thermally captured by the deep level defect during the electrical fill pulse and subsequently photoemitted to the valence band during illumination. DLOS for the negative $\Delta C(t)$ was not resolved well enough to determine $E^o$ accurately for hole photoemission from the deep level, but $E^o$ can be approximated as $h\nu$ where $\Delta C$ (and thus $N_t$) becomes negative, placing the corresponding deep level at approximately $E_v$ + 0.9 eV. The fact that the $E_v$ + 0.9 eV and $E_c$ – 1.06 eV deep level energies sum approximately to the band gap energy and that the transition between their corresponding negative and positive SSPC features is so abrupt suggests that these two deep levels are the same deep level with $e^{o,p} >> e^{o,n}$ for $h\nu < 1.10$ eV and $e^{o,n} >> e^{o,p}$ for $h\nu > 1.15$ eV. This $E_c$ – 1.06 eV/$E_v$ + 0.9 eV deep level is likely to be an effective recombination center because it can thermally capture carriers from both the conduction and valence band during the electrical fill pulse to produce both electron and hole photoemission. That is, this level is normally unoccupied and can efficiently capture both electrons and holes that have been generated either electrically or optically. The concentration of the $E_c$ – 1.06 eV/$E_v$ + 0.9 eV deep level was calculated as the change in net $N_t$ from $h\nu = 1.15$ eV (when the level is most occupied with electrons) to $h\nu = 1.50$ eV (when the level is least occupied).

Comparing the DLOS/SSPC spectra for *n*-InGaP photodetectors studied here and previous DLOS [33] and deep level transient spectroscopy (DLTS) [19] studies of *p*-InGaP base regions of solar cells provides insight into possible atomistic sources of the observed deep levels. Previous DLOS and DLTS of as-grown, *p*-InGaP [19,33] did not observe deep levels consistent with those reported here for *n*-InGaP. However, proton irradiation of *p*-InGaP produced a deep level at $E_v$ + 0.9 eV that was found to readily capture carriers from both the conduction and valence band [19]. It is concluded that the *p*-InGaP $E_v$ + 0.9 eV and the *n*-InGaP $E_c$ – 1.06/$E_v$ + 0.9 eV deep levels are likely the same defect based on their similar energy and ability to thermally interact with both the valence and conduction band. Thermal annealing investigations attributed the *p*-InGaP $E_v$ + 0.9 eV deep level to a phosphorus vacancy or a related anti-site [19].

DLOS and SSPC spectra were measured for the AlInGaP sample to compare to that of InGaP and are shown in Figs. 4 and 5, respectively. The AlInGaP has a band gap of 2.04 eV, as determined through photoluminescence. The DLOS spectrum for AlInGaP has qualitative similarities to that of InGaP, including a near-$E_v$ deep level at $E_c$ – 1.97 eV (fit using the Lucovsky model [30]), a deep level in the lower third of the band gap at $E_c$ – 1.80 eV with $d_{FC}$ = 0.43 eV (fit using the Pässler model [31]), and a broad line-shape around the middle of the band gap energy. It is noted that InGaP and AlInGaP DLOS spectra both exhibit deep levels with $E^o$ near 1.8 eV, however, these deep levels are not likely related. The InGaP $E_c$ – 1.81 eV deep level with $d_{FC}$ = 0 eV has a sharp $\sigma^o$ consistent with an effective-mass-like state with weak lattice coupling, while the AlInGaP $E_c$ – 1.80 eV deep level with $d_{FC}$ = 0.43 eV has a broad $\sigma^o$ consistent with a highly localized state with strong lattice coupling. Non-linear least-squares regression fitting of the DLOS data in Fig. 6 to the model of Pässler [31] produced an RMSE of 0.02 eV. This value was taken to be the uncertainty for both $E^o$ and $d_{FC}$ values for the $E_c$ – 1.80 eV deep level. Non-linear least-squares regression fitting of the DLOS data in Fig. 6 to the model of Lucovsky produced an RMSE = 0.01 eV, which was taken to be the uncertainty for $E^o$ for the $E_c$ – 1.97 eV deep level. The AlInGaP $E_c$ – 1.97 and the InGaP $E_c$ – 1.81 eV deep levels have nearly the same energy relative to their respective $E_v$, which suggests that they might have a similar atomistic source whose electronic structure is mainly derived from the valence band and thus tends to track $E_v$ with alloying [34].

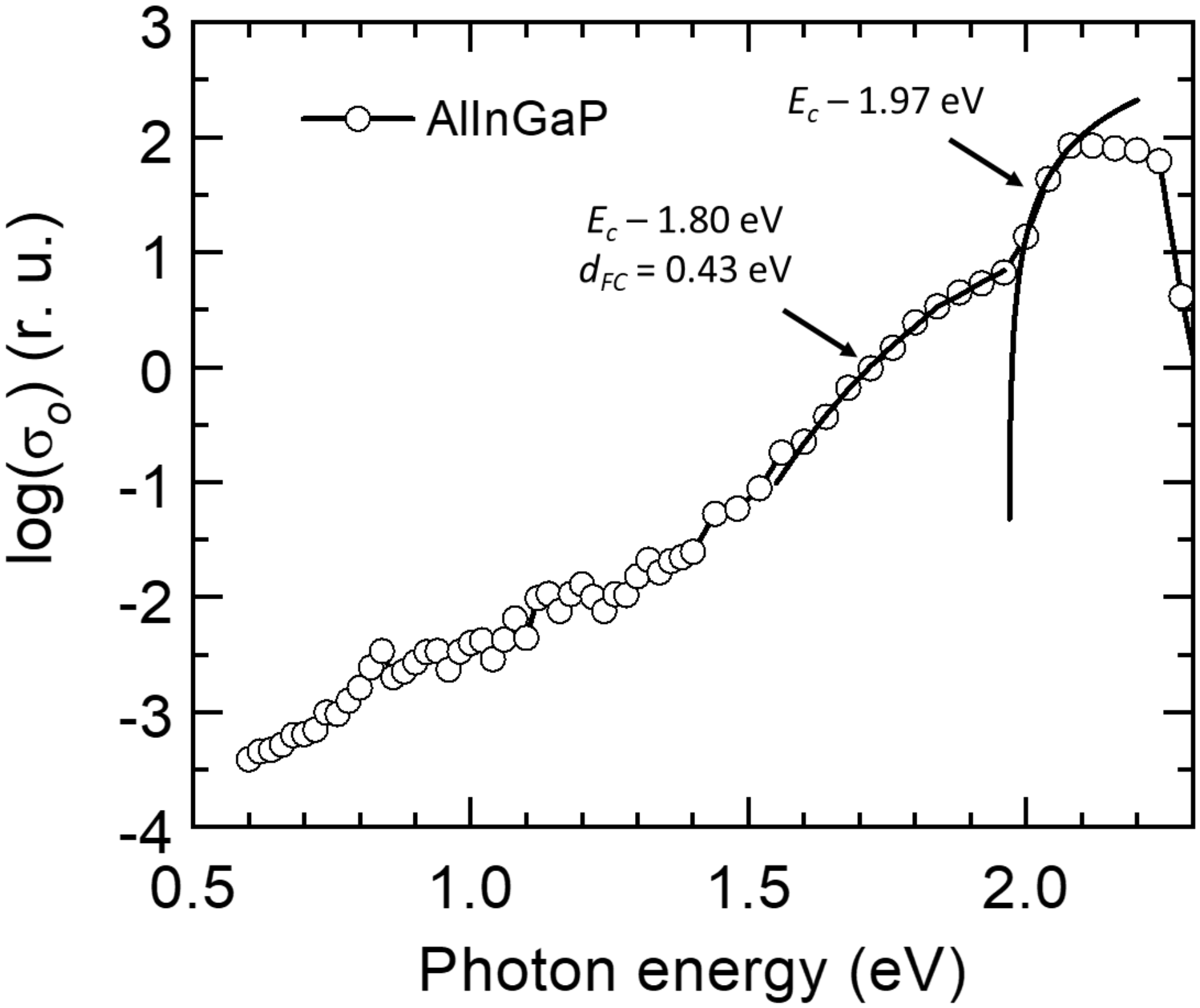

Figure 6. DLOS spectrum of the AlInGaP diode. The symbols are experimental data and the curves are fits to the Pässler model [31] for the $E_c$ – 1.80 eV level and the Lucovsky model [30] for the $E_c$ – 1.97 eV level. The line-shape at 0.60 – 1.50 eV was too broad to be well fit.

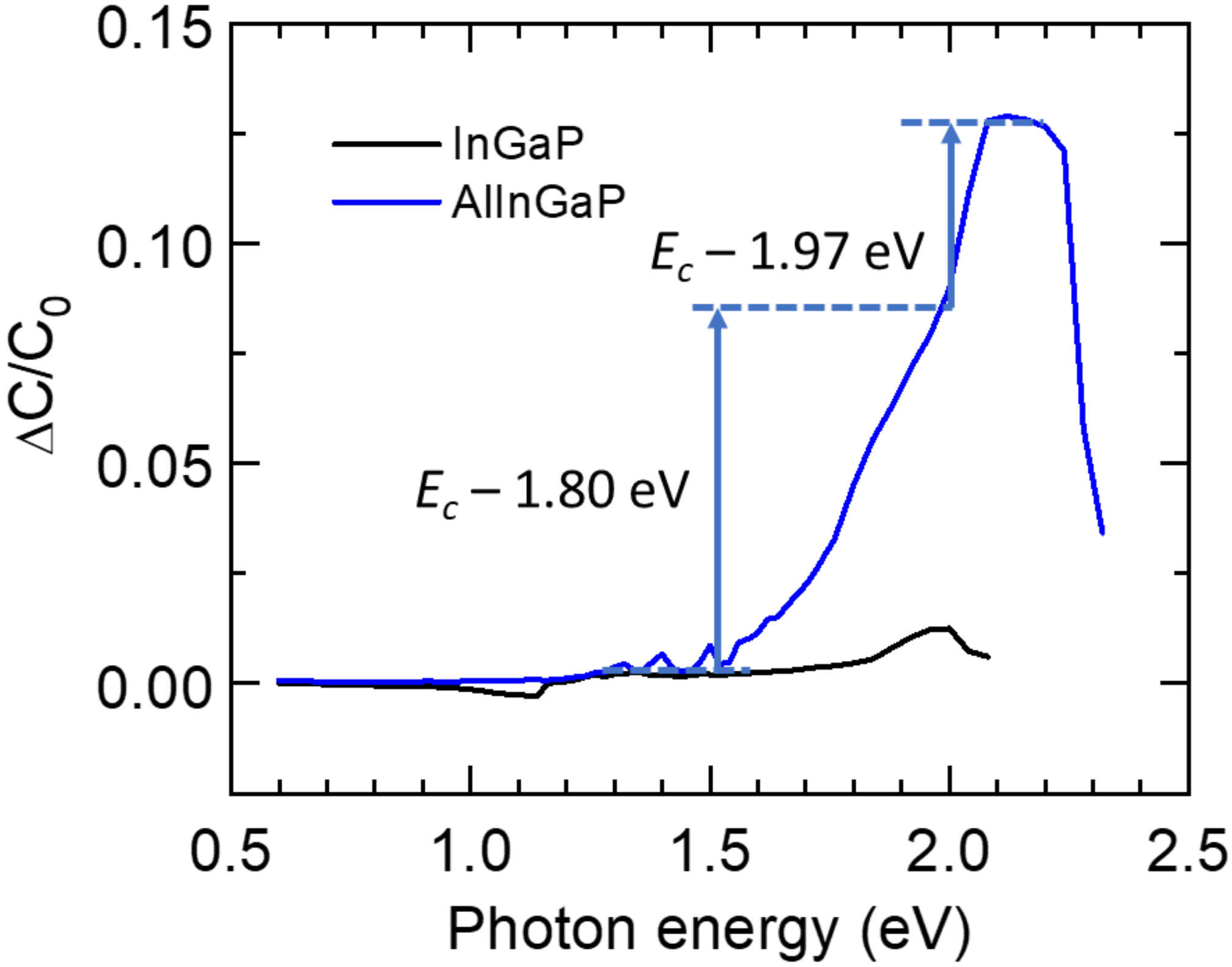

Figure 7. SSPC spectra of the InGaP and AlInGaP diodes. The InGaP data correspond to those in Fig. 4 but are recast as the relative change in photocapacitance. The arrows indicate $\Delta C/C_0$ of individual deep levels for the AlInGaP diode.

For defect states energetically located far from a band edge, the vacuum-referred binding energy (VRBE) model has been used among related compound semiconductors [34] and their alloys [35] to assess if they share a common atomistic source. In the VRBE model, a defect within a family of semiconductors is expected to form at an energy level $E_{vac}$ - $\chi$ - $E_{th}$, where $E_{vac}$ is the vacuum level, $\chi$ is the electron affinity and $E_{th} = E^o - d_{FC}$ is the thermodynamic energy level. The electron affinity for InGaP and AlInP alloys latticed-matched to GaAs only differ by approximately 0.1 eV [36], so it is reasonable to assume that $\chi$ for the alloys studied here are similarly close. In this case, comparing $E_{th}$ is a good proxy for the VRBE model. The AlInGaP $E_c$

– 1.80 eV deep level has $E_{th}$ = 1.37 eV and the InGaP $E_c$ – 1.58 eV level has $E_{th}$ = 1.42 eV. Thus, attributing these deep levels to the same atomistic source is consistent with the VRBE model, however, the atomistic source remains unknown. The difference in $d_{FC}$ for the $E_c$ – 1.80 eV and $E_c$ – 1.58 eV deep levels accounts for how their $E_{th}$ values can be similar despite very different $E^o$ values. The large $d_{FC}$ for the $E_c$ – 1.80 eV deep level indicates that, upon photoemission of a carrier, the change in local electronic bonding around the corresponding AlInGaP defect center induces a drastic spatial reconfiguration of the nearby atoms that releases substantial mechanical energy via phonon emission. The smaller $d_{FC}$ for the $E_c$ – 1.58 eV InGaP deep level suggests that the atomic reconfiguration around the corresponding defect center in InGaP is much less drastic and energetic after electron photoemission. One possible explanation of this scenario is that inclusion of Al in the matrix leads to a more "crowded" atomic configuration around the defect center and thus a more "rigid" coupling between local atoms and thereby a more energetic relaxation process when the local bonding is disrupted upon delocalization of a bound electron through photoemission.

The $\sigma^o$ line-shape for AlInGaP around the middle of the band gap indicates the presence of a near-mid-gap deep level. However, this line-shape was too broad to be well fit with the Pässler model, so $E^o$ and $d_{FC}$ could not be determined. Nonetheless, SSPC reveals significant differences between the near-mid-gap deep levels of the InGaP and AlInGaP diodes.

The SSPC spectra in Fig. 7 reveal two stark differences in the deep level spectra of InGaP and AlInGaP. First, the AlInGaP near-mid-gap deep level does not exhibit hole photoemission, i.e. negative $\Delta C$. This observation implies that the thermal hole capture cross-section for the AlInGaP near-mid-gap deep level is much smaller than the InGaP $E_c$ – 1.05 eV deep level. A much lower thermal hole capture cross-section for the mid-band gap deep level in the AlInGaP versus InGaP alloy would make the former a much less effective non-radiative recombination center. The second major difference revealed by the SSPC spectra in Fig. 7 is that the AlInGaP has a much larger overall $\Delta C/C_0$, i.e. $N_t/N_d$, compared to the InGaP diode. Indeed, $\Delta C/C_0$ for the AlInGaP diode was so large that LCV was used to determine $N_t$. The spatial profiles for the AlInGaP deep levels are shown in Fig. 8. Table I lists $N_t$ data for the deep levels of both alloy compositions. Since $n_{h\nu}$ were approximately constant with $x_d$, the typical $\sigma_{Nt}/N_t$ values for the data in Fig. 9 were 1.2, 1.3 and $1.5 \times 10^{15}$ cm$^{-3}$ for the near-mid-gap, $E_c - 1.80$ eV and $E_c - 1.97$ eV deep levels, respectively. LCV was not performed for the InGaP diode because its ~ $10^{15}$ cm$^{-3}$ uncertainty in $N_t$ was larger than $N_t$ values ~ $10^{14}$ cm$^{-3}$ determined from SSPC. Comparing these data to the net dopant concentration $N_d^* = N_d - \Sigma N_t$ measured by C-V (assuming all observed deep levels are acceptor-like) shows that the InGaP diode was subject to only 4% compensation compared to 25% compensation for the AlInGaP diode. Thus, the AlInGaP alloy poses a doping challenge due to a high deep level concentration but might nonetheless have longer minority carrier lifetime, and thus lower dark current, due to reduced thermal hole capture by mid-gap deep levels compared to the InGaP alloy.

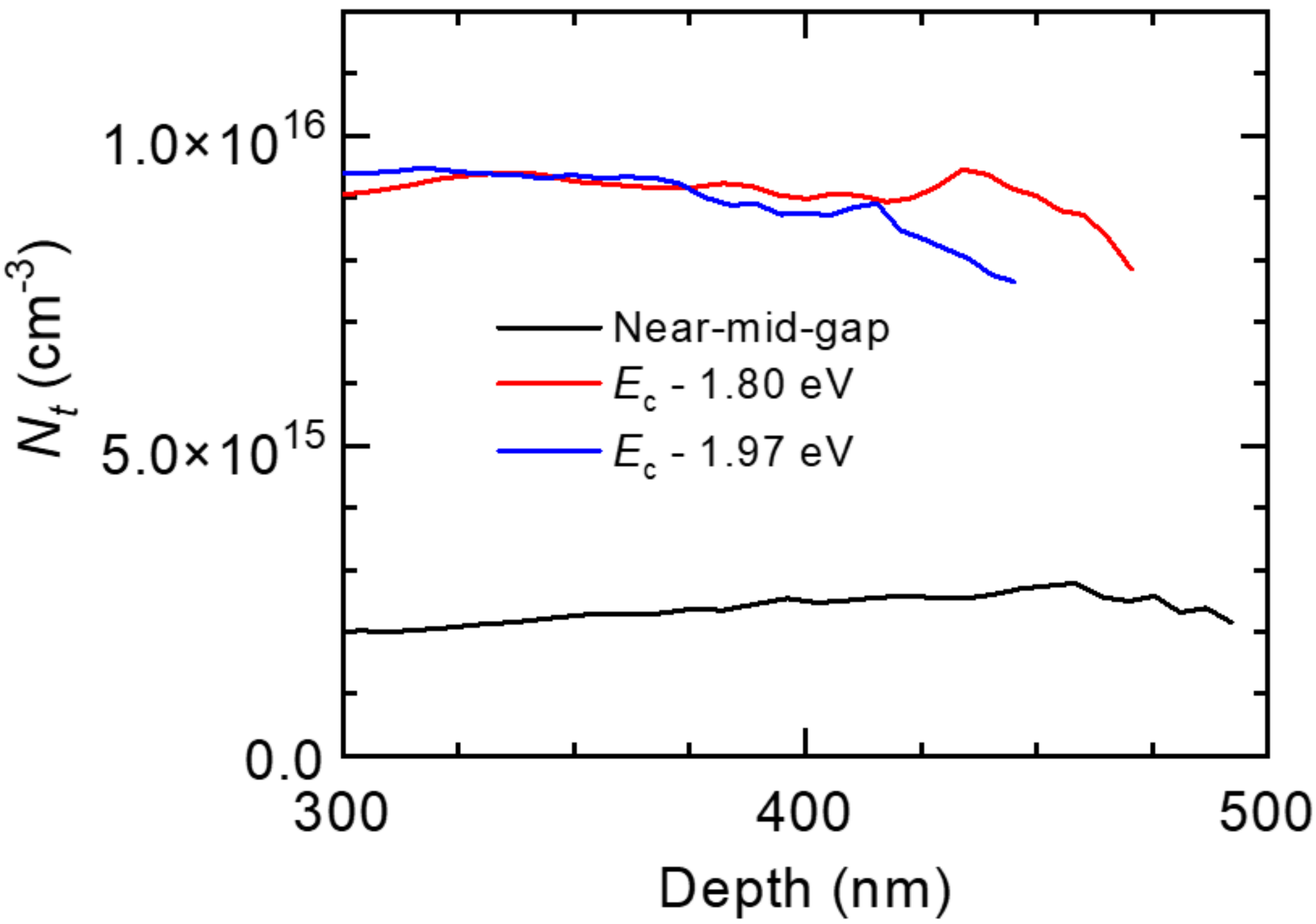

Figure 8. LCV data for the AlInGaP diode showing $N_t$ profiles determined from differencing the space-charge profiles shown in the inset of Figure 9.

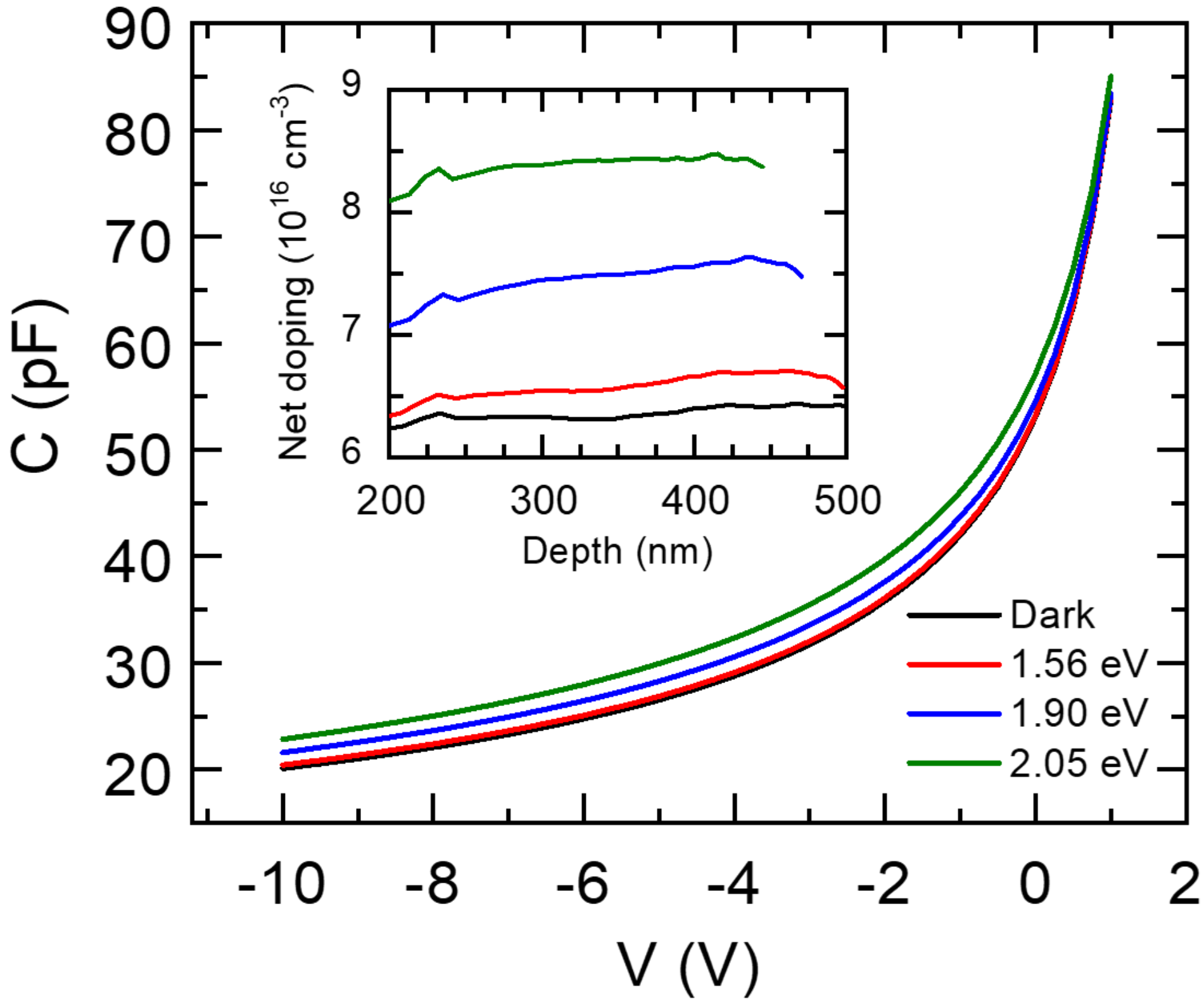

Figure 9. LCV data for the AlInGaP photodiode and extracted space-charge profiles under illumination (inset).

Table I. Deep level concentrations for both alloys.

| InGaP | | AlInGaP | |
|---|---|---|---|
| Energy (eV) | $N_t$ (cm$^{-3}$) | Energy (eV) | $N_t$ (cm$^{-3}$) |
| $E_c$ – 1.06/ $E_v$ + 0.9 | $4.0\times10^{14}$ | Near-mid-gap | $2.3\times10^{15}$ |
| $E_c$ – 1.58 | $1.5\times10^{14}$ | $E_c$ – 1.80 | $9.3\times10^{15}$ |
| $E_c$ – 1.81 | $8.2\times10^{14}$ | $E_c$ – 1.97 | $9.3\times10^{15}$ |

## 4. Conclusions

We have examined the defect characteristics of lightly doped *n*-type InGaP and AlInGaP alloys through several capacitance-based measurement techniques. Our analysis reveals the presence of several deep-level traps, with one in particular near midgap. The deepest optically identified defect energies in each material, $E_c$ – 1.97 eV and $E_c$ – 1.80 eV for AlInGaP and $E_c$ – 1.81 eV and $E_c$ –1.58 eV for InGaP, have similar thermodynamic energies and distances from band edges, suggesting analogous, though not identified, atomistic origins for these defects between the two materials. The inclusion of Al increases the density of the observed trap states by approximately an order of magnitude, however, the midgap state appears to have a reduced hole capture cross section in AlInGaP compared to InGaP and may thus be less efficient at mediating the generation of dark current. Our results suggest that the InGaP and AlInGaP materials may be good candidates for the development of SiPM-like devices optimized for shorter wavelength and lower dark count rate scintillator systems.

**Data Availability Statement**

The data that support the findings of this study are available from the corresponding author upon reasonable request.

DLOS capacitive transient fit available at https://doi.org/10.1088/1361-6641/ae1b37/data1.

**Acknowledgements**

We gratefully acknowledge helpful conversations with Jeffrey A. Ivie. This work was supported by the Laboratory Directed Research and Development Program at Sandia National Laboratories. Sandia National Laboratories is a multi-mission laboratory managed and operated by National Technology & Engineering Solutions of Sandia, LLC (NTESS), a wholly owned subsidiary of Honeywell International Inc., for the U.S. Department of Energy's National Nuclear Security Administration (DOE/NNSA) under contract DE-NA0003525. This written work is authored by an employee of NTESS. The employee, not NTESS, owns the right, title and interest in and to the written work and is responsible for its contents. Any subjective views or opinions that might be expressed in the written work do not necessarily represent the views of the U.S.

Government. The publisher acknowledges that the U.S. Government retains a non-exclusive, paid-up, irrevocable, world-wide license to publish or reproduce the published form of this written work or allow others to do so, for U.S. Government purposes. The DOE will provide public access to results of federally sponsored research in accordance with the DOE Public Access Plan.